\documentclass[reprint,amsmath,amssymb,pra,groupedaddress,nobibnotes,showpacs]{revtex4-1}
\usepackage{graphicx}

\begin{document}

\title{Storage and Retrieval of Thermal Light in Warm Atomic Vapor}

\author{Young-Wook Cho}
\email{choyoungwook81@gmail.com}
\affiliation{Department of Physics, Pohang University of Science and Technology (POSTECH), Pohang, 790-784, Korea}

\author{Yoon-Ho Kim}
\email{yoonho72@gmail.com}
\affiliation{Department of Physics, Pohang University of Science and Technology (POSTECH), Pohang, 790-784, Korea}

\date{\today}

\begin{abstract}
We report slowed propagation and storage and retrieval of thermal light in warm rubidium vapor using the effect of electromagnetically-induced transparency (EIT). We first demonstrate slowed-propagation of the probe thermal light beam through an EIT medium by measuring the second-order correlation function of the light field using the Hanbury-Brown$-$Twiss interferometer. We also report an experimental study on the effect of the EIT slow-light medium on the  temporal coherence of thermal light. Finally, we demonstrate the storage and retrieval of thermal light beam in the EIT medium. The direct measurement of the photon number statistics of the retrieved light field shows  that the photon number statistics is preserved during the storage and retrieval process.
\end{abstract}

\pacs{32.80.Qk, 42.50.Gy, 42.50.Ar}

\maketitle


Actively controlling the speed of a light pulse has been extensively studied in the recent years using the phenomenon known as electromagnetically-induced transparency (EIT). In the EIT effect, a strong laser field (the ``coupling beam'') creates a narrow transparency window for the probe field within an otherwise, i.e., without the coupling field, opaque region of the atomic spectra as a result of induced quantum interference among the interacting atomic energy levels \cite{eit}. The peculiarity of the EIT effect is that the newly opened transparency window can exhibit extraordinarily steep spectral dispersion, potentially resulting in an ultra-slow group velocity for the probe pulse \cite{hau,kash}.

Furthermore, the group velocity of the probe field can even be reduced to zero, effectively freezing the probe pulse within the atomic coherence medium, by adiabatically changing the coupling field amplitude so that the quantum state of light is coherently transferred to the atomic spin excitation in a completely reversible manner \cite{dark,dark2}. So far, EIT-based light storage has been demonstrated for a coherent light pulse \cite{Lukin}, a single-photon \cite{single1,single2}, frequency filtered spontaneous parametric down-conversion photons \cite{SPDC,SPDC2}, and the squeezed vacuum \cite{squeezed,honda}.

To date, however, slowed propagation and storage and retrieval of (chaotic) thermal light in an EIT medium have not been demonstrated. Since chaotic light can mimic certain features of quantum correlations, e.g., $N$-photon ghost interference and imaging \cite{scarcelli,Valencia,aga,cao,chen09b}, and can be used for  chaotic communication \cite{chaotic}, it is of importance and interest to consider how chaotic light would interact with an EIT medium and if the chaotic nature of thermal light would indeed be preserved during the storage and retrieval process.

In this paper, we report what we believe to be the first experimental demonstration of slow propagation and storage and retrieval of thermal light in an EIT medium. First, we demonstrate EIT-induced slowed-propagation of the probe thermal light beam in a warm atomic (Rubidium 85) vapor by measuring the second-order correlation function of the light field using the Hanbury-Brown$-$Twiss interferometer. We also study how the temporal coherence of the thermal light is affected by the EIT slow-light medium. We then demonstrate the storage and retrieval of a thermal light beam in the EIT medium. In particular, by directly measuring the photon number statistics of the retrieved light beam, we show that the photon number statistics is preserved during the storage and retrieval process.

\begin{figure}[b]
\centering
\includegraphics[width=3in]{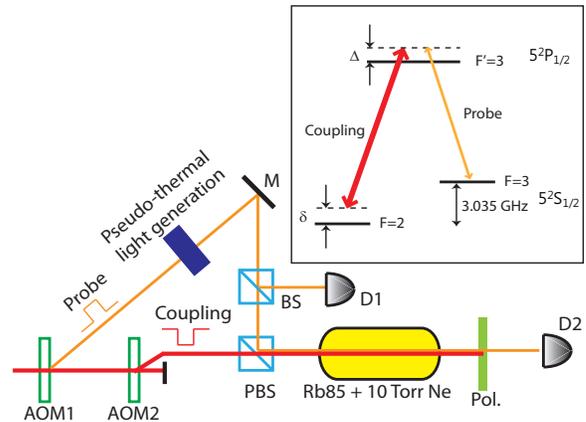}
\caption{Schematic diagram of the experiment. The generation of pseudo-thermal probe beam is realized with an electro-optic modulator driven by a Gaussian noise voltage or by scattering a focused laser beam at a rotating ground disk. The inset shows the the energy level configuration of \textsuperscript {85}Rb D1 transition line used in this experiment. AOM: acousto-optic modulator, BS: non-polarizing beam splitter, PBS: polarizing beam splitter, and Pol.: Glan-Thompson polarizer. D1 and D2 are detectors.} \label{setup}
\end{figure}


Consider the experimental setup shown in Fig.~\ref{setup}. An external cavity diode laser (ECDL) operating at 795 nm is used as the main light source for this experiment. First, the laser beam from ECDL is sent to an acousto-optic modulator (AOM1) in the double pass configuration to generate an additional laser beam which is 2.956 GHz down-shifted with respect to the undiffracted main beam. The main beam is then sent to a second acousto-optic modulator (AOM2) which generates a new laser beam, 80 MHz up-shifted with respect to the main beam. The laser beams generated at AOM1 and AOM2 are 3.035 GHz apart which is equal to the hyperfine splitting of two ground states of \textsuperscript{85}Rb. The laser beam generated at AOM2 is used as the (strong) coupling beam and the laser beam generated at AOM1 is used as the (weak) probe beam. Note that the coupling beam is horizontally polarized and the probe beam is vertically polarized (waveplates are not shown in the figure). The probe laser beam is then used to generate (chaotic) thermal light either by passing it through an electro-optic modulator driven by a Gaussian noise voltage \cite{Xie} or by scattering the focused laser beam at a rotating ground disk \cite{thermal,are}.

The coupling and probe fields are combined at PBS and directed to a 75 mm long quartz cell filled with isotopically pure \textsuperscript{85}Rb vapor with 10 Torr Ne buffer gas. A heating element surrounds the vapor cell and the heater-cell package was wrapped with $\mu$-metal shields to reduce the effect of stray magnetic fields. The vapor cell was heated to maintain 50$\sim$55$^{\circ}$C during the experiment. The coupling field is then blocked by a polarizer (Pol) and only the probe field is detected at D2. After the polarizer, the power ratio of the residual coupling beam to the probe beam is less than 0.3\% in our experiment.

Our $\Lambda$-type EIT scheme makes use of the \textsuperscript{85}Rb D1 line as shown in the inset of Fig.~\ref{setup}. The probe field couples the  $5^{2}S_{1/2}$ $F=3$ $\rightarrow$ $5^{2}P_{1/2}$ $F'=3$ transition with the frequency detuning  of $\Delta\approx100$ MHz to the blue in order to achieve better transmission and the coupling field is set to satisfy the two-photon resonance condition $\delta=0$.

Before doing the thermal light storage experiment, we first measured the EIT bandwidth for a probe laser beam by detuning the frequency of the coupling laser. The full width at half maximum (FWHM) EIT window was measured to be 200 kHz. The frequency of the coupling laser was set so that the probe laser transmission was at its maximum, i.e., the coupling and the probe fields satisfied the two-photon resonance condition.


Let us first describe the slow-light experiment for thermal light. First, to generate a (chaotic) thermal probe field, the probe beam generated at AOM1 is sent through an electro-optic modulator (EOM) driven by a Gaussian noise voltage with its intensity and phase randomly set \cite{Xie}. The noise bandwidth is varied from 40 kHz to 500 kHz depending on a particular experimental run. The average power of the thermal probe field and the laser coupling field were 50 $\mu$W and 3 mW, respectively.

The normalized second-order correlation function, $g^{(2)}$, is measured with the Hanbury-Brown$-$Twiss interferometer to confirm the thermal light characteristics of the probe field and to observe the slowed-propagation of the thermal light via the EIT medium. The Hanbury-Brown$-$Twiss interferometer is formed with a 50/50 beam splitter BS and two photocurrent detectors D1 and D2 (Thorlabs, model PDB150A), see Fig.~\ref{setup} \cite{hbt}. The cross-correlation of the intensity fluctuations between the two detectors is given as  \cite{loudon}
\begin{equation}
\left\langle {\Delta I_1 (t_1 )\Delta I_2 (t_2 )} \right\rangle   \propto g^{(2)} \left(\tau \right) - 1,
\end{equation}
where $\Delta I_i(t_i)$ is the AC component of the current measured at detector $i$ in time $t_i$, $\tau= t_2  - t_1$, and $\langle \ldots \rangle$ represents the time average. In experiment, the photocurrent is first digitized and the $\tau$-dependent cross-correlation function is evaluated  on a computer by using the AC components of the signals.

\begin{figure}[t]
\centering
\includegraphics[width=3in]{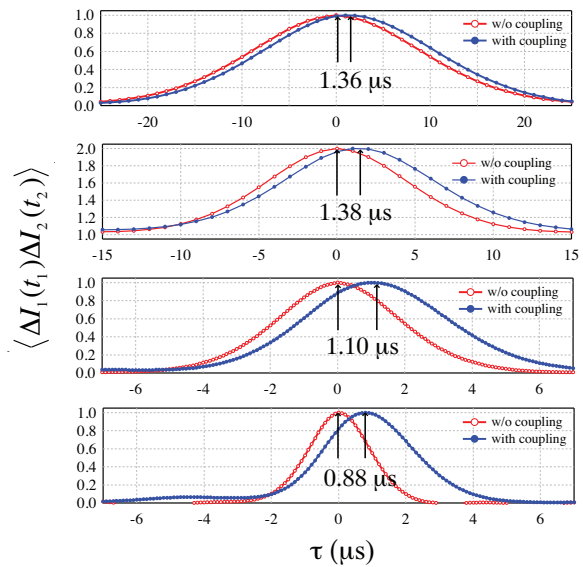}
\caption{Slowed propagation of chaotic light in the EIT medium. The normalized cross-correlation function gets broadened when the incident thermal light bandwidth $\sigma$ is sufficiently larger than the 200 kHz EIT bandwidth. 
The measured bandwidths $\sigma$ of the transmitted thermal light in the EIT condition (i.e., with the coupling on) are (a) 46 kHz, (b) 80 kHz, (c) 202 kHz and (d) 322 kHz. All plots are normalized to unity.}\label{slow} 
\end{figure}

The slowed-propagation of thermal light is demonstrated by observing the second-order correlation of the light field with or without the coupling beam. For (chaotic) thermal light with a Gaussian spectral width $\sigma$, the normalized second-order correlation function takes the form of a Gaussian function with the temporal coherence time corresponding to $\sigma$ \cite{loudon}. If we, therefore, assume that the net effect of the EIT medium to the chaotic light is just a group delay, the normalized second-order correlation function for the slowed thermal light is given as
\begin{equation}
g^{(2)} (\tau ) -1 = \exp [ - \sigma ^2 (\tau  - \tau _g )^2 \ln 2],
\end{equation}
where $\tau_g$ is the delay due to the slow-light condition of the EIT medium in the presence of the coupling field. 


Figure \ref{slow} shows the experimental data for slowed-propagation of thermal light in the EIT medium. The experimental data clearly show the effect of the EIT medium to the chaotic light. When the bandwidth of the probe thermal light is much less than the 200 kHz EIT bandwidth, the second-order correlation function is delayed with no change in its shape, see Fig.~\ref{slow}(a). However, as inferred from Fig.~\ref{slow}(b), (c) and (d), the second-order correlation function gets broadened as the bandwidth of the thermal light is increased. The broadening of $g^{(2)}$ is caused by the fact that, when the bandwidth of the probe thermal light becomes larger than the EIT bandwidth, the probe beam gets spectrally filtered due to frequency-dependent absorption and higher-order dispersion also comes into play. Note also that, as the bandwidth of the thermal light gets larger than the EIT bandwidth, the amount of delay is apparently reduced, see Fig.~\ref{slow}(c) and Fig.~\ref{slow}(d).


We now describe the experiment on the storage and retrieval of thermal light in the EIT medium. The experimental schematic is similar to Fig.~\ref{setup} but with small changes. First, the chaotic light for this measurement was generated by focusing the probe laser beam (diffracted at AOM1) on a rotating ground disk rather than using the Gaussian noise driven EOM as done in the previous slow-light experiment. It is because the statistical properties of the EOM-driven chaotic light depend entirely on the characteristics of the driving electronic signal. Thus, unwanted electronic noises, such as the power noise, may easily alter the measured photon count distribution. On the other hand, the chaotic light generated by scattering a focused laser beam on a rotating ground disk has been known to offer an excellent Bose-Einstein photon count distribution \cite{thermal,are}. In this experiment, the peak power of the probe thermal light pulse and the coupling laser pulse were 10 $\mu$W and 3 mW, respectively. The spectral bandwidth of the thermal probe was measured to be $\sigma=104$ kHz by using the Hanbury-Brown$-$Twiss interferometer described earlier. Second, the detector D2 is now replaced with a single-mode fiber coupled single-photon counting detector for photon count distribution measurement to directly observe the photon number statistics of the retrieved probe field in the EIT medium.

\begin{figure}[t]
\centering
\includegraphics[width=3in]{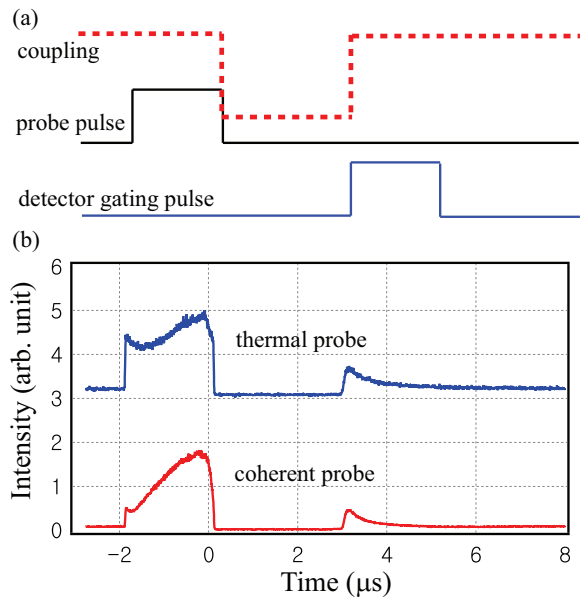}
\caption{(a) Pulse sequence for the light storage and retrieval experiment. The detector gating pulse for D2 is necessary for directly measuring the photon count distribution of the retrieved light. (b) Storage and retrieval of the chaotic thermal probe field and the coherent probe field. Note that the small kink at leading edge of leakage pulse (around $-2$ $\mu$s) is  just an overshoot due to the AOM driver response.  
}\label{storage}
\end{figure}

Figure \ref{storage}(a) shows the pulse sequence for the storage and retrieval of thermal light. The coupling laser beam is initially turned on and this prepares the EIT medium. The probe field (chaotic thermal light), shaped into a 2 $\mu$s rectangular pulse by turning on and off AOM1, is then directed to the EIT medium. The probe pulse gets compressed temporally as it enters the vapor cell due to the reduced group velocity in the EIT medium. After the probe pulse has completely entered the vapor cell, the coupling laser is turned off, mapping the state of the probe field onto the atomic states. After the 3 $\mu$s storage duration, the stored probe pulse is retrieved by turning back on the coupling field. The repetition rate of the whole sequence was  8 kHz. 
 
The experimental results are shown in Fig.~\ref{storage}(b). For this measurement, we use a time-correlated single-photon counter (TCSPC) in the start-stop histogram mode with the 8 kHz synchronization pulse as the start and the D2 signal as the stop. The use of the single-mode fiber and a moderate neutral-density filter ensures that D2 has one photon count event on average per measurement cycle. The storage and retrieval of the chaotic probe field as well as the coherent probe field are demonstrated in Fig.~\ref{storage}(b). The experimental result clearly demonstrates that chaotic thermal light can be dynamically stored and retrieved in an EIT medium.

Although the experimental results in Fig.~\ref{storage}  show that chaotic thermal light can be stored and retrieved in warm atomic vapor using the EIT effect, it is necessary to perform the photon number statistics measurement to see if the retrieved field indeed exhibits the photon count distribution of the chaotic thermal light. The photon count distribution can be measured by using a photoelectric detector and the probability of registering $n$ count events during the measurement interval $T$ is given as \cite{Mandel}
 \begin{equation}
P(n) = \left\langle {\frac{{(\eta \overline {\mathcal{I}} (t,T)T)^n }}{{n!}}\exp [ - \eta \overline  {\mathcal{I}} (t,T)T]} \right\rangle,\label{pn}
\end{equation}
where $\eta$ is the overall detection efficiency, $\overline  {\mathcal{I}} (t,T)$ is the average intensity of the light field during the measurement time $T$, and $\langle \ldots \rangle$ represents the time average. For the coherent state of light, eq.~(\ref{pn}) is calculated to be the Poisson distribution \cite{thermal,are,loudon}
\begin{equation}
P_{coh}(n) = \frac{ {\bar n} ^n }{n!} \exp [ - {\bar n} ],
\end{equation}
where ${\bar n} = \eta \overline  {\mathcal{I}} (t,T) T$ is the average  value of photon counts for the measurement interval $T$. For chaotic light, eq.~(\ref{pn})  reduces to the Bose-Einstein distribution \cite{thermal,are,loudon}
\begin{equation}
P_{th}(n) = \frac{{\bar n} ^n }{{\left( {1 + {\bar n} } \right)^{1 + n} }}.
\end{equation}

We have performed an experiment to directly measure the photon number statistics of the retrieved field.
Since it is essential to measure the photon count events only for the retrieved light field, a 2 $\mu$s electronic gating pulse, which sets the measurement interval $T$, was applied to the detector D2 right after the coupling field was turned back on, see Fig.~\ref{storage}(a). The single-photon counting detector we used, Perlkin-Elmer SPCM-AQ4C, has roughly 50 ns dead time and 400 Hz dark count rate. Note that the $T=2$ $\mu$s gate pulse duration was chosen so that it is much shorter than the coherence time of the probe field while much larger than the dead time of the detector.

\begin{figure}[t]
\centering
\includegraphics[width=3in]{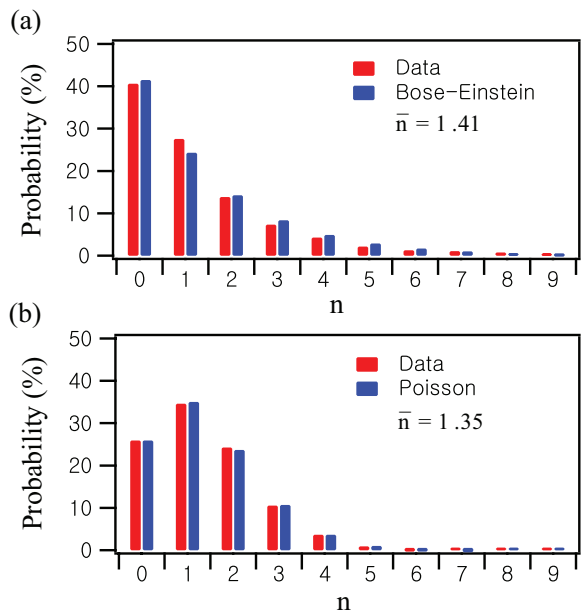}
\caption{Measured and calculated photon count distribution $P(n)$ for the retrieved probe field. (a) The probe field is a chaotic light generated at the rotating ground disk. (b) The probe field is a coherent laser beam.  }\label{stat}
\end{figure}

The experimental results of the photon counting distribution measurement for the retrieved probe field from the EIT medium are shown in Fig.~\ref{stat}. As shown in Fig.~\ref{stat}(a), for the case that the probe field is chaotic thermal light, the measured photon number statistics for the retrieved field follows the Bose-Einstein distribution. When the probe field is coherent (the ground disk is not rotated), the retrieved field exhibits the Poisson distribution, see Fig.~\ref{stat}(b).

If we approximate that the probe field as the single mode field, the normalized second-order correlation  coefficient is given as \cite{loudon}
\begin{equation}
g^{(2)}=1+\left({(\Delta n)^{2}-\bar n}\right)/ {{\bar n}^{2}},
\end{equation}
where $(\Delta n)^{2}=\langle n^2 \rangle - \bar n ^2$ and $\bar n$ can be calculated from the experimental data. For Fig.~\ref{stat}(a), we get $g^{(2)}=2.23\pm0.132$ and for Fig.~\ref{stat}(b), we get $g^{(2)}=0.995\pm0.016$. These values agree quite well with the theoretical values:  $g^{(2)}=1$ for coherent light and $g^{(2)}=2$ for thermal light. Potential sources of errors for the experimental values are issues related to the single-photon detector (dark counts and dead time) and the fact that the probe field is in fact not single-mode. (We did not directly measure $g^{(2)}(\tau)$ of the retrieved field due to electronics limitations, see Ref.~\cite{note}.)


In summary, we have experimentally investigated propagation and storage of chaotic thermal light in an EIT medium. We have demonstrated the slow light effect for the thermal light in the EIT medium by observing the delayed cross-correlation function using a Hanbury-Brown$-$Twiss interferometer. We have also demonstrated storage and retrieval of chaotic thermal light in the EIT medium and showed that the photon count distribution is preserved during the storage and retrieval process.

We would like to thank H. Kang and H.-S. Moon for helpful comments. This work was supported by the National Research Foundation of Korea (2009-0070668 and 2009-0084473) and POSTECH BSRI Research Fund. YWC acknowledges partial support from the Korea Research Foundation (KRF-2008-314-C00075).



\begin{thebibliography}{99}
\bibitem{note} To measure $g^{(2)}(\tau)$ of the stored/retrieved field, it is essential to gate the detectors so that the leakage part of the pulse is not measured. For example, in Fig.~\ref{slow}, the probe fields measured for $-2$ $\mu$s $\sim  0$ $\mu$s correspond to the leakage. The single-photon detector (Perkin-Elmer SPCM-AQ4C) we used, however, always produces one spurious TTL output pulse per gate pulse when operated in the gated mode. Since measuring $g^{(2)}(\tau)$ involves coincidence measurements between two single-photon detectors (both of which to be operated in the gated mode), the spurious TTL output of the SPCM in the gated mode becomes a serious problem for directly measuring $g^{(2)}(\tau)$. Also, the TCSPC we used (PicoHarp 300) cannot be gated.
\end{thebibliography}
\end{document}